\documentstyle[aps,preprint,epsf]{revtex}

\newcommand{\ds}[1]{#1 \hspace{-0.5em}/}  
\newcommand\beq{\begin{equation}}
\newcommand\eeq{\end{equation}}
\newcommand\beqa{\begin{eqnarray}}
\newcommand\eeqa{\end{eqnarray}}
\newcommand\bzeta{\mbox{\boldmath$\zeta$}}
\newcommand\bp{{\bf k}}
\newcommand\bq{{\bf q}}
\newcommand\E{\epsilon}

\begin{document}

\textwidth=16.0cm
\textheight=23.5cm 

\draft 
 
\title{\hfill{\normalsize KUNS- 1611}\\
Ferromagnetism of quark liquid}
\author{Toshitaka Tatsumi\\
Department of Physics, Kyoto University\\
Kyoto 606-8502, Japan}

\maketitle
 
\begin{abstract}
Usually it is believed that the Hartree-Fock state of quark matter is
a Fermi gas state with no polarisation of spins. We examine the
possibility of the polarised quark liquid interacting with the
one-gluon-exchange interaction. It is suggested that the 
Hartree-Fock state shows a spontaneous magnetic instability at low
densities through the same mechanism as the appearance of 
ferromagnetism in electron gas. Metastability of the polarised
quark liquid is also discussed. 

\noindent PACS: 04.40.Dg, 07.55.Db, 12.38.Bx, 26.60.+c, 97.60.Jd

\noindent Keywords: neutron stars, ferromagnetism, quark matter,
magnetic field

\end{abstract}
\newpage

The physical origin of the strong magnetic field in neutron stars is a
long-standing problem \cite{cha}. Recent discovery 
of ``magnetars'' \cite{tho,ob} 
- neutron stars with
magnetic field strengths in excess of $B_{cr}=m_e^2c^3/e\hbar\sim
4.4\times 10^{13}$G and typically $\sim 10^{15}$G -
also stimulate our interest in
the origin of their strong magnetic field. It
has been widely believed that the magnetic field of neutron stars 
is inherited from the progenitor stars \cite{cha}. 
However, the extraordinary magnetic field 
in magnetars seems to enforce our reconsideration of its origin. 
Inside neutron stars, hadronic matter exists beyond the nuclear density
($n_0\simeq 0.16$fm$^{-3}$), 
so that it may be interesting to
consider the hadronic origin of the magnetic field.

In this Letter we examine the possibility of the
spin-polarised quark liquid by the use of the Fermi liquid theory,
where quarks interact on each other by 
the one-gluon-exchange (OGE) interaction. 
Since the
system should be totally color singlet (neutral), there is left only the
exchange energy within the Hartree-Fock approximation.
It has been suggested that three-flavor quark matter (strange quark
matter) can be stable or metastable for a reasonable range of QCD
parameters \cite{chi,wit,far,mad}. 
The appearance of strange ($s$) quarks is favored at and
above the nuclear density due to the reduction of the kinetic energy
of the system 
and the specific feature of OGE interaction; the exchange energy is
negative for massive strange quarks at low densities, 
while it is always positive for massless $u$ and $d$ quarks. This
feature is very similar to the Coulomb interaction for electron
gas. In the following we consider the possibility of 
ferromagnetism in strange quark matter on the analogy with 
electron gas.

In 1929 Bloch first pointed out the possibility of ferromagnetism of
electron gas at low density \cite{blo}: electrons interact on each other by
the exchange Coulomb force, which contribution is attractive and   
becomes dominant over the
kinetic energy at low density due to their different density
dependence. Then the unpolarised electron gas suddenly 
turns into the completely polarised state at the critical density (the
first order phase transition). The critical density is determined by
the mass of electrons and the QED coupling constant $\alpha=e^2/4\pi$; 
lower density, where the nonrelativistic approximation works very well, 
is favored for the spontaneous magnetisation.
This idea has been confirmed
theoretically by the use of the Green function Monte-Calro method \cite{cep}.

Therefore if the exchange
energy is attractive and dominant over the kinetic energy at some 
density, there should appear ferromagnetism also in strange quark matter, 
as in electron gas.  
The idea can be put into a
relativistic formulation utilizing the polarisation density matrix \cite{ber}. 
Then we shall see a difference between theories of electron gas and quark
liquid. The scattering amplitude of two quarks 
becomes momentum dependent, and includes not only the
spin-nonflip term but also the spin-flip one due to the lower
component of the Dirac spinor, the latter of which
vanishes in the non-relativistic limit. Also, OGE interaction itself 
include the
spin-dependent (magnetic)   
interaction as well, which becomes effective in the relativistic
regime, besides the Coulomb-like (electric) interaction.
These features somewhat change
the mechanism of polarisation in quark liquid from in electron
gas. Actually we shall see that there is a possibility of spontaneous
magnetisation even in the relativistic regime, where the exchange
energy contribution is no longer attractive, by a different reason
from electron gas.

Consider the spin-polarised quark liquid with the total number density 
of quarks $n_q$
\footnote{We,hereafter, consider one flavor quark matter, since OGE
interaction never mix flavors.}
; we denote the number densities of 
quarks with spin up and down by $n_+$ and $n_-$, respectively
\footnote{The meaning of spin up and down should be carefully defined
in the relativistic theories. Here we call them in the rest frame of each 
particle. Their precise meaning is given later.}
, and introduce the
polarisation parameter $p$ by the
equations,
\beq
n_+=\frac{1}{2}n_q(1+p), \qquad n_-=\frac{1}{2}n_q(1-p),
\label{dd}
\eeq
under the condition $0\leq p\leq 1$. We assume as usual 
that three color states 
are occupied to be neutral for each momentum and spin state.
The Fermi momentum in the
spin-polarised quark matter then is 
\beq
k_F^+=k_F(1+p)^{1/3}, \quad k_F^-=k_F(1-p)^{1/3}
\label{de}
\eeq
with the Fermi momentum for the unpolarised matter, $k_F=(\pi^2n_q)^{1/3}$.
The kinetic energy density is given by the standard formula,
\beq
\E_{kin}=\frac{3}{16\pi^2}\sum_{i=\pm}\left[k_F^iE_F^i(2k_F^{i2}+m^2_q)
-m^4_q\ln\left(\frac{E_F^i+k_F^i}{m_q}\right)\right],
\label{fa}
\eeq
with the Fermi energy $E_F^i=(m^2_q+k_F^{i2})^{1/2}$. 

The interaction energy is given by the OGE interaction.
There is no direct energy contribution
(Hartree term) and left only the exchange energy contribution (Fock term).
To characterize the degeneracy of the plane wave solution 
$u^{(\alpha)}(k)$ ($\alpha=1,2$) for 
a positive energy state, we introduce a space-like 4-pseudovector
$a^\mu$ which in the rest frame is twice the three-dimensional mean spin
vector $\bzeta$ with $|\bzeta|=1$, $a^\mu=(0, \bzeta)$. The 4-vector
in any frame, where the particle  
is moving with the momentum $\bp$, is
found by a Lorentz transformation from the rest frame, and 
\beq
{\bf a}=\bzeta+\frac{\bp(\bzeta\cdot\bp)}{m_q(E_k+m_q)}, 
~a^0=\frac{\bp\cdot\bzeta}{m_q}
\label{ac}
\eeq
with $E_k=(m^2_q+|\bp|^2)^{1/2}$. By choosing $\bzeta$ along the $z$
axis, $\bzeta\equiv (0,0,\pm 1)$, we see that the spinors
$u^{(\alpha)}(k)$ are eigenstates of the operator $-W\cdot a/m_q$, where 
the 4-vector $W^\mu$ is the Pauli-Lubanski vector, 
$W^\mu=-\frac{1}{4}\epsilon_{\mu\nu\rho\sigma}k^\nu\sigma^{\rho\sigma}$.
For the standard
representation of $u^{(\alpha)}(k)$ \cite{itz}, the eigenvalue is 
$+1/2$(spin-up) for $u^{(1)}(k)$, and $-1/2$(spin-down) for $u^{(2)}(k)$. 
Then the polarisation density matrix $\rho$ is given by the expression,
\beq
\rho(k, \zeta)
=\frac{1}{2m_q}(\ds{k}+m_q)P(a), \quad P(a)=\frac{1}{2}(1+\gamma_5\ds{a}),
\label{ad}
\eeq
which is normalized by the condition, ${\rm tr}\rho(k, \zeta)=1$.

Let us 
consider the interaction between two quarks with momenta, $k$ and $q$, and 
the different spin vectors, $\bzeta$ and $\bzeta'$, respectively.
They are specified by the polarisation density matrices, 
$\rho(k,\zeta)$ and $\rho(q,\zeta')$. 
The Landau Fermi-liquid interaction $f_{{\bf k}\zeta,{\bf q}\zeta'}$
is related to the two-particle forward scattering amplitude \cite{bay},
\beq
 f_{{\bf k}\zeta,{\bf q}\zeta'}
=\frac{m_q}{E_k}\frac{m_q}{E_q}{\cal M}_{{\bf k}\zeta,{\bf q}\zeta'}
\label{ce}
\eeq
where ${\cal M}_{{\bf k}\zeta,{\bf q}\zeta'}$ is the usual Lorentz
invariant matrix element. The color symmetric matrix element is given
by the exchange term only; the direct term vanishes because the color
symmetric combinations ($\sim {\rm tr}\lambda_a$) does not couple to
gluons. Thus    
\beqa
{\cal M}^s_{{\bf k}\zeta,{\bf q}\zeta'}&=&-g^2
\frac{1}{9}{\rm tr}(\lambda_a/2\lambda_a/2)\bar u^{(\zeta')}(\bq)\gamma_\mu
u^{(\zeta)}(\bp)\bar u^{(\zeta)}(\bp)\gamma^\mu u^{(\zeta')}(\bq)
\frac{-1}{(k-q)^2}
\nonumber\\
&=&\frac{4}{9}g^2\frac{1}{4}{\rm tr}
\left[\gamma_\mu\rho(k,\zeta)\gamma^\mu\rho(q,\zeta')\right]
\frac{1}{(k-q)^2}.
\label{cf}
\eeqa
Substituting Eqs.~(\ref{ac}) and (\ref{ad}) into Eq.~(\ref{cf}), we
finally find 
\beqa
{\cal M}^s_{{\bf k}\zeta,{\bf q}\zeta'}
&=&\frac{2}{9}g^2\left[2m^2_q-k\cdot q
-(\bp\cdot\bzeta)(\bq\cdot\bzeta')
+m^2_q\bzeta\cdot\bzeta'\right.\nonumber\\
&+&\left.\frac{1}{(E_k+m_q)(E_q+m_q)}
\left\{m_q(E_k+m_q)(\bzeta\cdot\bq)(\bzeta'\cdot\bq)
+m_q(E_q+m_q)(\bzeta'\cdot\bp)(\bzeta\cdot\bp)\right.\right.\nonumber\\
&+&\left.\left.(\bp\cdot\bq)(\bzeta\cdot\bp)
(\bzeta'\cdot\bq)\right\}\right]\frac{1}{(k-q)^2}.
\label{cg}
\eeqa
after some manipulation of $\gamma$-matrix algebra. If we choose both
$\bzeta$ and $\bzeta'$ along the $z$ axis in parallel, 
$\bzeta=\bzeta'=(0,0,\pm 1)$, we have the spin-nonflip amplitude 
${\cal M}^{s, nonflip}_{\bp\bq}$, while if we choose them in anti-parallel,
$\bzeta=-\bzeta'=(0,0,\pm 1)$, we have the spin-flip amplitude 
${\cal M}^{s, flip}_{\bp\bq}$. Each form of the spin-nonflip or
spin-flip amplitude is complicated, but the average of them gives a
simple form,
\beq
\overline{{\cal M}}^s_{\bp\bq}=\frac{2}{9}g^2\frac{2m^2_q-k\cdot q}{(k-q)^2},
\label{cj}
\eeq
which is nothing but the matrix element for the unpolarised case \cite{bay}.
In the nonrelativistic limit, $m_q\gg |\bp|, |\bq|$, the matrix
element is reduced into the form,
\beq
{\cal M}^s_{{\bf k}\zeta,{\bf
q}\zeta'}=\frac{2}{9}g^2\frac{m_q^2(1+\bzeta\cdot \bzeta')}{(k-q)^2}, 
\eeq
so that there is {\it no correlation} between quarks with different
spins. On the other hand, there is some correlation included even in
the Hartree-Fock approximation for the relativistic case.

After summing up over the
color degree of freedom and performing the integrals of the
Fermi-liquid interactions, $f_{\bp\bq}^{s, nonflip}= 
(m_q/E_k)(m_q/E_q){\cal M}^{s, nonflip}_{\bp\bq}$ and 
$f_{\bp\bq}^{s, flip}= 
(m_q/E_k)(m_q/E_q){\cal M}^{s, flip}_{\bp\bq}$, over the Fermi
seas, we have the exchange energy
density $\E_{ex}$ consisting of two contributions,
\beq
\E_{ex}=\E_{ex}^{nonflip}+\E_{ex}^{flip}.
\label{da}
\eeq
The spin-nonflip contribution can be written as
\beq
\E_{ex}^{nonflip}=\frac{\alpha_c}{4\pi^5}(I(k_F^+)+I(k_F^-)),
\label{eo}
\eeq
where $\alpha_c$ is the fine structure constant of QCD,
$\alpha_c=g^2/4\pi$. The integral $I(k_F^i) (i=\pm)$ is given by 
\beq
I(k_F^i)=\frac{\pi^2}{2}\xi_i\left(\xi_i-\frac{4m_q}{3}\kappa_i\right)
+\tilde I(k_F^i), 
\label{eq}
\eeq
with $\xi_i=E_F^ik_F^i-m^2_q\ln(E_F^i+k_F^i/m_q)$ and 
$\kappa_i=k_F^i-m_q\ln(E_F^i+k_F^i/m_q)$,
and the remaining integral
\beq
\tilde I(k_F^i)=\frac{2\pi^2m_q}{3}\left[\int_0^{k_F^i}qdq
\frac{(E_q+m_q)(E_q+2m_q)+m^2_q}{E_q(E_q+m_q)}K_1(q,k_F^i)
-\int_0^{k_F^i}\frac{qdq}{E_q+m_q}K_2(q,k_F^i)\right],
\label{za}
\eeq
with the functions,
\beq
K_1(q,k_F^i)=(E_F^i-E_q)
\ln\left|\frac{m^2_q-E_F^iE_q+k_F^iq}{m^2_q-E_F^iE_q-k_F^iq}\right|
-2q\ln\left(\frac{E_F^i+k_F^i}{m_q}\right)
\eeq
and
\beq
K_2(q,k_F^i)=\frac{1}{2}\left[(k_F^{i2}-q^2)
\ln\left|\frac{m^2_q-E_F^iE_q+k_F^iq}{m^2_q-E_F^iE_q-k_F^iq}\right|
+qE_q\ln\left(\frac{E_F^i-k_F^i}{E_F^i+k_F^i}\right)
-2k_F^iq\right]
\eeq
Similarly the spin-flip contribution can be written as 
\beq
\E_{ex}^{flip}=2\frac{\alpha_c}{4\pi^5}J(k_F^+, k_F^-).
\label{ep}
\eeq
The integral $J(k_F^+, k_F^-)$ is given by 
\beq
J(k_F^+, k_F^-)=\frac{\pi^2}{2}\left[\xi_+\xi_-
+\frac{2m_q}{3}\left(\xi_+\kappa_-+\xi_-\kappa_+\right)\right]
+\tilde J(k_F^+, k_F^-), 
\label{er}
\eeq
where 
\beqa
\tilde J(k_F^+, k_F^-)&=&-\frac{\pi^2m_q}{3}
\left[\int_0^{k_F^-}\frac{qdq}{E_q+m_q}(E_qK_1(q,k_F^+)
-K_2(q,k_F^+))\right.\nonumber\\
&+&\left.\int_0^{k_F^+}\frac{pdp}{E_p+m_q}(E_pK_1(p,k_F^-)-K_2(p,k_F^-))\right].
\eeqa

For the unpolarised ($p=0$) matter the sum of $I$
and $J$ in Eqs.~(\ref{eq}) and (\ref{er}) gives 
\beq
\E_{ex}^{unpol}\equiv 2\frac{\alpha_c}{4\pi^5}(I+J)=-\frac{\alpha_c}{\pi^3}
\left[k_F^4
-\frac{3}{2}
\left\{E_Fk_F-m^2_q\ln\left(\frac{E_F+k_F}{m_q}\right)\right\}^2
\right]
\label{et}
\eeq
with the Fermi energy $E_F$ for the unpolarised case,
$E_F=(m^2_q+k_F^2)^{1/2}$ \cite{bay,akh}.

In the nonrelativistic case, $k_F^i\ll m_q$, the kinetic energy density
is written as
\beq
\E_{kin}\sim\frac{3k_F^5}{20\pi^2m_q}
\left\{(1+p)^{5/3}+(1-p)^{5/3}\right\}.
\label{fb}
\eeq
The spin-flip contribution becomes tiny in this case, and 
the dominant contribution for the OGE energy density in
Eq.(\ref{da}) comes from the spin-nonflip contribution, especially
$\tilde I$ given in Eq.~(\ref{za}), $\tilde I\sim -2\pi^2k_F^{i4}$.
Thus we find an attractive contribution,
\beq
\E_{ex}\sim -\frac{\alpha_ck_F^4}{2\pi^3}
\left\{(1+p)^{4/3}+(1-p)^{4/3}\right\}.
\label{fd}
\eeq
The exchange energy contribution is attractive and exactly
given by the Coulomb-like (spin-nonflip) interaction of gluons. There is no 
spin-flip contribution in the nonrelativistic limit. The form of the 
energy densities (\ref{fb}) and (\ref{fd}) is the same as in electron gas.
It is the difference of
density dependence between these contributions given 
in Eqs.~(\ref{fb}) and (\ref{fd}) which may cause a
ferromagnetic instability: 
as density becomes low enough, all the particles suddenly 
align their spins and the completely
polarised ($p=1$) matter is energetically favorable. This mechanism is firstly
discovered by Bloch for electron gas \cite{blo}; in
the charge neutral system, electrons tend to avoid the Coulomb
repulsion from each other. If all electrons have the same spin state, they can
maximally avoid the repulsion by the Pauli principle. On the other
hand, the kinetic energy of electrons obviously becomes minimum for the
unpolarised ($p=0$) case. We can see 
that there is a trade-off between two contributions.    
Thus the critical density is determined by the mass and the coupling constant.

In the relativistic case there are some different features from the
nonrelativistic case. First, there is a spin-flip
contribution due to the lower component of the Dirac spinor even for
the Coulomb-like interaction. Secondly, the transverse (magnetic) gluons becomes
important, where the spin-flip effect is prominent. Finally, 
the density dependence of kinetic energy as
well as the exchange energy is very different \cite{tam}.
Before discussing the general case, we consider the relativistic
limit, $k_F^i\gg m_q$; the kinetic energy density behaves like
\beq
\E_{kin}\sim \frac{3k_F^4}{8\pi^2}\left\{(1+p)^{4/3}+(1-p)^{4/3}\right\}.
\label{fe}
\eeq
In this limit, $I(k_F^i)\sim \pi^2/2k_F^{i4}$ and 
$J(k_F^+,k_F^-)\sim \pi^2/2k_F^{+2}k_F^{-2}$. Thus 
the exchange energy looks like 
\beq
\E_{ex}\sim \frac{\alpha_c}{8\pi^3}k_F^4
\left\{(1+p)^{4/3}+(1-p)^{4/3}+2(1-p^2)^{2/3}\right\}.
\label{ff} 
\eeq
The kinetic energy is simply an increasing function with respect to 
the polarisation parameter $p$ and takes a minimum at $p=0$, 
whereas the exchange energy is the decreasing function and takes
a minimum at $p=1$. This is due to the characteristic feature 
of the spin-flip and the
spin-nonflip interactions. Both give a repulsive contribution in the
relativistic limit. In the polarised state ($p=1$) there is no spin-flip
interaction and the spin-nonflip contribution remains unchanged 
in spite of the increase of the Fermi momentum.
The difference of the total energy, $
\E_{tot}=\E_{kin}+\E_{pot}$, at $p=0$ and $p=1$ reads
\beq
\delta \E\equiv \E_{tot}(p=1)-\E_{tot}(p=0)
=\frac{3k_F^4}{8\pi^2}(2^{4/3}-1)\left[1-\frac{4-2^{4/3}}{2^{4/3}-1}
\frac{\alpha_c}{\pi}\right].
\label{fz}
\eeq
Hence if $\alpha_c>3.23$ the polarised state is energetically favorable.
Thus, ferromagnetism in the relativistic limit has a defferent
origin from
that in the nonrelativistic case: the spin-flip contribution favors
the polarised state.

In Fig.1 a typical shape of the energy density is depicted 
as a function of the polarisation parameter $p$, e.g. for the set
,$m_q=300$MeV of the $s$ quark and 
$\alpha_c=2.2$ as in the MIT bag model \cite{deg}
\footnote{The difficulties to determine the values of these parameters 
have been discussed in ref. \cite{far}, and we must allow some range
for them.}.
\begin{center}
\fbox{\thicklines\parbox{0.06\textwidth}
{ Fig.1
}}
\end{center}
For small $p\ll 1$, the energy density behaves like
\beq
\E_{tot}-\E_{tot}(p=0)=\chi^{-1} p^2+O(p^4)
\label{ha}
\eeq
with $\chi^{-1}\equiv \chi_{kin}^{-1}+\chi_{pot}^{-1}$. $\chi$ is 
proportional to the magnetic susceptibility, and its sign change
suggests a ferromagnetic transition. It consists of two contributions: 
the kinetic energy gives $\chi_{kin}^{-1}=k_F^5/(3\pi^2E_F)$ (the
Pauli paramagnetism), which 
changes from $\chi_{kin}^{-1}\sim O(k_F^5)$ at low densities to
$\chi_{kin}^{-1}\sim O(k_F^4)$ at high densities.
On the other hand, the exchange energy gives
\beq
\chi_{pot}^{-1}=-\frac{2\alpha_ck_F^4}{9\pi^3}\left[2-\frac{6k_F^2}{E_F^2}
+3\left\{E_Fk_F-m_q^2\ln\left(\frac{E_F+k_F}{m_q}\right)\right\}
\frac{k_F}{E_F^3}
+2\frac{k_F^2}{E_F^2}\left\{1+\frac{2m_q}{3(E_F+m_q)}\right\}\right].
\label{hc}
\eeq
In
the relativistic limit, $p_F\gg m_q$, $\chi_{pot}^{-1}$ behaves like 
\beq
\chi_{pot}^{-1}\sim
\frac{\alpha_c}{9\pi^3}k_F^4-\frac{\alpha_c}{3\pi^3}k_F^4
=-\frac{2\alpha_c}{9\pi^3}k_F^4
\label{hd}
\eeq
from Eq.~(\ref{ff}), where the first term stems from the spin-nonflip
contribution, while the second term from the spin-flip contribution.
Then we can see that the effect of the spin-flip contribution
overwhelms the one of the spin-nonflip contribution. 
The potential contribution
$\chi_{pot}^{-1}$ is always {\it negative}
, and dominant over $\chi_{kin}^{-1}$ 
at low densities,
while the kinetic contribution $\chi_{kin}$ is always {\it
positive}. If $\alpha_c>3\pi/2=4.7$, $\chi$ becomes negative over all
densities.

For a given set of $m_q$ and $\alpha_c$, $\chi$ changes its sign 
at the critical density, denoted by $n_{c1}$, and it 
clearly shows that quark liquid is ferromagnetic below 
that density. This is equivalent with the sign change of the curvature 
of the free energy at the origin as a function of the polarisation
parameter $p$. Note that the ferromagnetic
transition in our case is of the first order, so that it is not
sufficient to only see the susceptibility. Even above that density the
ferromagnetic phase may be
possible. Actually there is a range, $n_{c1}<n_q<n_{c2}$, 
which is specified by
the condition  s.t. $\chi>0$ and $\E<0$.
Above the density $n_{c2}$ there is no longer the stable
ferromagnetic phase. However, the {\it metastable} state is still possible
up to the density $n_{c3}$, which is specified by the condition
s.t. $\eta\equiv\partial \E_{tot}/\partial p~|_{p=1}<0$. In Fig.2 we
depict the quantities $\chi^{-1},\delta \E$  and $\eta$ as functions of
density, e.g. for the set
,$m_q=300$MeV and $\alpha_c=2.2$.
The crossing points with the horizontal axis indicate the critical
deisities $n_{c1}, n_{c2}$ and $n_{c3}$, respectively. 
\begin{center}
\fbox{\thicklines\parbox{0.06\textwidth}
{ Fig.2
}}
\end{center}
We can see that the ferromagnetic instability occurs at low densities, 
while the metastable state can exist up to rather high densities.

Finally we show the critical lines satisfying $\chi^{-1}=0, \delta \E=0$ and 
$\eta=0$
in the QCD parameter ($\alpha_c$ and $m_q$) plane, which
seperate the three characteristic regions  
for a given density. In Fig.~3 they are
presented at a density $n_q=0.3$fm$^{-3}$. All the lines have the
maxima at the intermediate quark mass, and the mechanism of  
ferromagnetism is different for each side of 
the maximum, as already discussed. 
If we take $m_q=300$MeV for the $s$ quark or $m_q\sim 0$MeV for the
$u$ or $d$ quark, and 
$\alpha_c=2.2$ as in the MIT bag model again \cite{deg}, the 
quark liquid can be
ferromagnetic as a metastable state.
\begin{center}
\fbox{\thicklines\parbox{0.06\textwidth}
{ Fig.3
}}
\end{center}

In this Letter we have pointed out a possibility of ferromagnetism of
quark liquid: quark liquid potentially bears a nature to be
ferromagnetic due to the OGE interaction. We have calculated the
energy density of the spin polarised quark liquid interacting with the 
OGE interaction in a relativistic framework, and found that the spin-flip and
spin-nonflip contributions exhibit an interesting feature as a function of
the polarisation parameter $p$ or density. The ferromagnetic transition 
in quark liquid is of the 
first order as in electron gas and there is a metastable state even
above the critical density. 

We have found that ferromagnetic instability is feasible not only
in massive quark system but also in light quark system. The mechanism is
different between two systems: the similar mechanism to electron gas
can be applied to the former system, while a novel mechanism 
due to the large spin-flip contribution,
which arises as a specific feature of the relativistic system,
may be applied to the latter system.

We have seen that the ferromagnetic phase is realized at low densities 
and the metastable state is plausible up to rather high densities for
a reasonable range of the QCD parameters. Our calculation is based on
the Hartree-Fock approximation. So we need to examine the higher-order
gluon-exchange 
contributions, especially the ring diagrams, which has been known to
be important in the calculation of the susceptibility 
of electron gas \cite{cep,bru}. The ring diagrams have been calculated 
for the unpolarised quark matter \cite{bay,kap}, 
but there is no calculation for the polarised matter.

The lifetime of the metastable state is another remaining problem to be
studied. It may be caused by the highly nonperturbative quantum effects. So 
it should be proportional to the exponential factor at least. Hence it must be
difficult to estimate the lifetime reliably.

If a ferromagnetic quark liquid exists stably or metastably around or
above nuclear density, it has
some implications on the properties of strange quark stars and strange
quark nuggets \cite{mad}. They should be magnetized in a macroscopic
scale. 
So strange quark nuggets may be no longer a candidate of dark
matter in the universe, since they can emit the electromagnetic waves 
like pulsars. 

For a consideration of 
a possibility to attribute magnetars to strange quark stars 
in a ferromagnetic phase,
we roughly estimate the strength of the magnetic field 
at the surface of a strange quark star. 
The stellar parameters of strange quark stars 
are similar to those for normal neutron stars
for the typical mass around $M_G=1.4M_\odot$ \cite{hae,alc}.
The magnetic dipole moment $M_q$ amounts to 
$M_q=\mu_q\cdot(4\pi/3\cdot r_q^3)n_q$ for the quark sphere with
the radius $r_q$, 
where $\mu_q$ is the magnetic moment of each quark, $\mu_q\sim
\mu_N$($\mu_N: $nuclear magneton$\sim 5\times 10^{-24}{\rm erg}\cdot
{\rm gauss}^{-1}$) for massive quarks and $10^2\mu_N$ for light
quarks. 
Then the dipolar magnetic field at
the star surface $r=R\simeq 10$Km takes a maximal strength at the poles, 
\beq
B_{max}=\frac{8\pi}{3}\left(\frac{r_q}{R}\right)^3\mu_qn_q,
\label{gc}
\eeq
which is order of $O(10^{15-17})$G for $n_q=O(0.1)$fm$^{-3}$, 
enough for the magnitude required for magnetars. 

One may wonder whether such a strong magnetic field modifies the Fermi 
sea of quark liquid itself. We can see, however, that the mean magnetic 
field working on one quark , which is produced by other quarks,  
vanishes due to the uniform distribution of quarks. Hence the
self-consistency is retained even in such a case. 
\vskip 1cm

The author thanks M. Matsuzaki and T. Muto for useful discussions 
and suggestions on the manuscript.
He also thanks N. Iwamoto for stimulating discussions.
This work was supported in part by the Japanese Grant-in-Aid for
Scientific Research Fund of the Ministry of Education, Science, Sports and
Culture (11640272).

\begin{figure}[h]
  \epsfsize=0.6\textwidth
\centerline{\epsffile{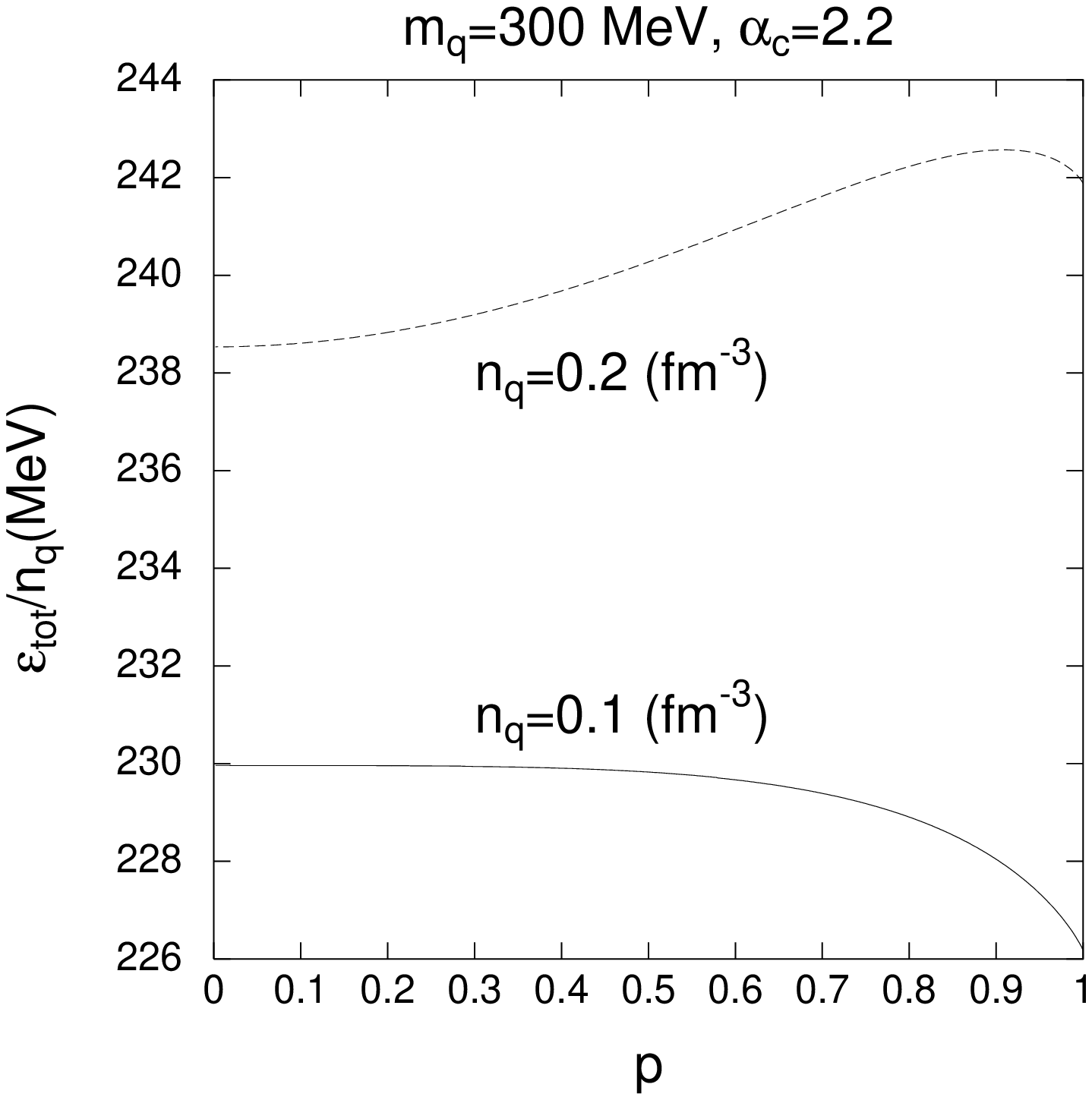}}
 \caption{Plot of the energy density as a function of the polarisation 
parameter at $n_q=0.1$fm$^{-3}$ and $n_q=0.2$fm$^{-3}$.}
 \label{fig1}
\end{figure}

\begin{figure}[h]
  \epsfsize=0.6\textwidth
\centerline{\epsffile{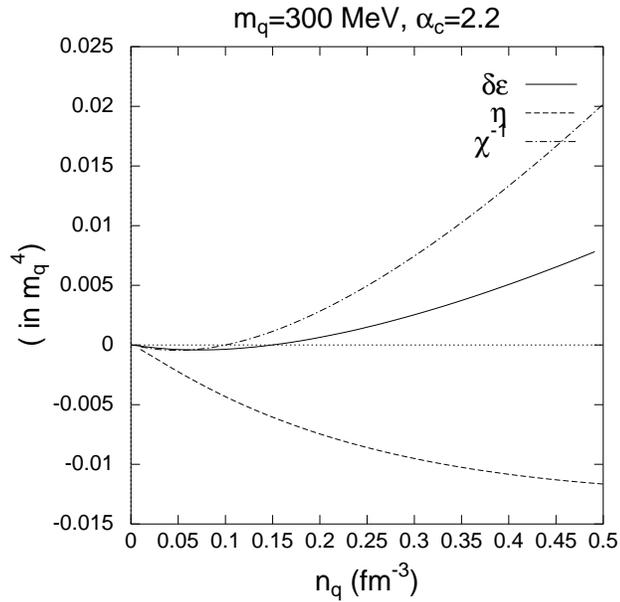}}
 \caption{Density dependence of $\chi^{-1},\delta \E$ and $\eta$. In this
case, $n_{c1}\simeq 0.1$fm$^{-3}$, 
$n_{c2}\simeq 0.15$fm$^{-3}$ and $n_{c3}>0.5$fm$^{-3}$.}
 \label{fig2}
\end{figure}

\begin{figure}[h]
  \epsfsize=0.6\textwidth
\centerline{\epsffile{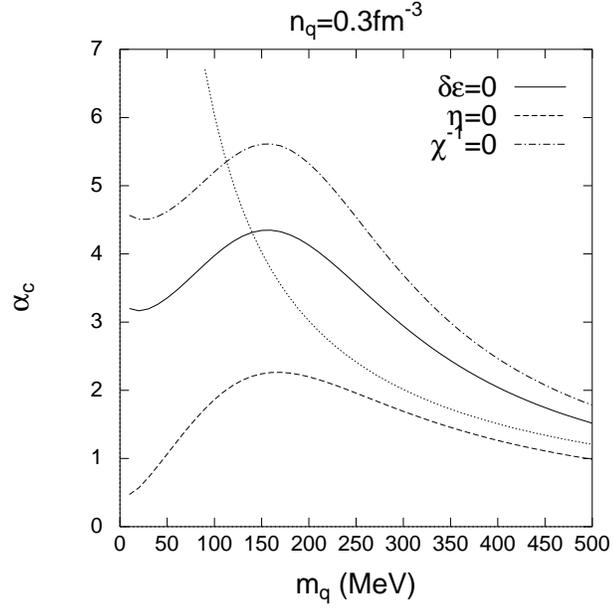}}
 \caption{Phase diagram in the QCD parameter plane. The region above
the solid curve is in a ferromagnetic phase. The region bounded by the 
solid and dased curves is in a paramagnetic phase, but a metastable
ferromagnetic state is possible. The dash-dotted curve shows the
boundary where the susceptibility is vanished. 
The dotted curve shows
the corresponding one to the solid curve with the nonrelativistic
formulae given in Eqs.~(21) and (22) for comparison.}
 \label{fig3}
\end{figure}

\end{document}